\documentclass[aps,floatfix,prd]{revtex4}

\usepackage{xspace}
\usepackage{graphicx}

\newcommand{\gev}  {\ensuremath{\mathrm{\;Ge\kern -0.1em V}}\xspace}
\newcommand{\gevc}{\ensuremath{{\mathrm{\,Ge\kern -0.1em V\!/}c}}\xspace}
\newcommand{\gevcc}{\ensuremath{{\mathrm{\,Ge\kern -0.1em V\!/}c^2}}\xspace}
\newcommand{\mev}  {\ensuremath{\mathrm{\;Me\kern -0.1em V}}\xspace}
\newcommand{\mevc}{\ensuremath{{\mathrm{\,Me\kern -0.1em V\!/}c}}\xspace}
\newcommand{\mevcc}{\ensuremath{{\mathrm{\,Me\kern -0.1em V\!/}c^2}}\xspace}
\newcommand{\ppbar}{\ensuremath{\mathrm{p\overline{p}}}\xspace}
\newcommand{\epem} {\ensuremath{\mathrm{e^+e^-}}\xspace}
\newcommand{\invfb}{\ensuremath{\mbox{\,fb}^{-1}}\xspace}

\begin{document}
\newcommand{\BABARPubYear}    {05}
\newcommand{\BABARPubNumber}  {071}
\newcommand{\SLACPubNumber} {11528}

\begin{flushleft}
hep-ex/0510042\\
BABAR-PROC-\BABARPubYear/\BABARPubNumber \\
SLAC-PUB-\SLACPubNumber
\end{flushleft}

\title{Studies of \epem Collisions with a Hard Initial-State Photon at BaBar \footnote{Work supported by Department of Energy contract DE-AC02-76SF00515.}}

\author{Nicolas Berger \\ Representing the BaBar Collaboration}
\affiliation{SLAC MS 61, Menlo Park, California, U.S.A.}

\begin{abstract}
We present preliminary BaBar measurements of hadronic cross sections in \epem annihilation 
using the radiative return technique. The cross sections for $\epem \to \ppbar\,$, 
 $\,3(\pi^+\pi^-)$, $\,2(\pi^+\pi^-)2\pi^0$, and $\,K^+K^-2(\pi^+\pi^-)$ are 
measured. Measurements of the proton form factor and of 
the ratio $G_E/G_M$ are also shown.
\end{abstract}

\maketitle

\section{Introduction}

In recent years, precise new results in low-energy hadronic cross sections
have been obtained at high-luminosity \epem machines using the radiative return 
technique. The cross section for $\epem \to \pi^+\pi^-$ has been measured by
the KLOE collaboration~\cite{KLOE}, while those of $\epem \to \pi^+\pi^-\pi^0$,
$\,\pi^+\pi^-\pi^+\pi^-$, $\,K^+K^-K^+K^-$ and
$\,K^+K^-\pi^+\pi^-$ have been measured by BaBar~\cite{threepi,fourh}. These
measurements are of particular interest since they provide input to data-driven
calculations
of hadronic contributions to the muon anomalous magnetic moment, $a_{\mu}^{had}$, and of
the running of the QED coupling constant $\Delta\alpha^{(5)}_{had}$, which appears
in global fits to the standard model. As a continuation of this program, 
preliminary BaBar measurements for the \ppbar, $\,3(\pi^+\pi^-)$, 
$\,2(\pi^+\pi^-)2\pi^0$, and $\,K^+K^-2(\pi^+\pi^-)$ final states
are presented here. The \ppbar and six-hadron results are based on $240 \invfb$ and
$232 \invfb$ of data respectively.

\section{Radiative Return}

The radiative return method uses the emission of an initial-state photon 
to probe center-of-mass energies $\sqrt{s'}$ below the
nominal collision energy $\sqrt{s}$. The emission of a photon of energy
$E_{\gamma}$ in the center-of-mass frame corresponds to the production of a 
recoiling system with $s' = s - 2\sqrt{s}\,E_{\gamma}$.
Cross-section measurements can therefore be obtained over a wide range of 
center-of-mass energies without varying the beam energy, thereby avoiding the usual 
point-to-point uncertainties associated with discrete measurements.

At BaBar the range $\sqrt{s'} \le 4.5 \gev$ is readily accessible. 
 To reduce backgrounds, events are reconstructed by requiring 
the ISR photon to be detected within the acceptance of the
BaBar electromagnetic calorimeter.
Since the recoiling system is produced as a jet opposite the high-energy photon,
the event detection efficiency is only weakly dependent on $\sqrt{s'}$, and 
measurements down to threshold are possible. The efficiency also does not depend strongly
on the simulation of the hadronic final state.

At first order the cross section for an ISR process with a photon polar angle acceptance of
$\theta_{min} < \theta_{\gamma} < \pi - \theta_{min}$ in the center-of-mass
frame is

\begin{equation}
\frac{d\sigma_{\epem \to \gamma X}}{dx} = \frac{\alpha}{\pi}\frac{1+x^2}{x(1-x)}
\left(\log\frac{1-\cos\theta_{min}}{1+\cos\theta_{min}} - (1-x)cos\theta_{min}\right)
\sigma_{\epem \to X}(s(1-x))
\end{equation}

\noindent where $x = 1 - s'/s$.
 A kinematic fit is used to measure $s'$ and reject
backgrounds. Cross sections are normalized using the process $\epem \to \mu^+\mu^-\gamma$.

\section{The \lowercase{$\,\ppbar\,$} final state}

The cross section for $\epem \to \ppbar$ is given by

\begin{figure}
\begin{minipage}{0.28\textwidth}
\includegraphics[width=1.0\textwidth]{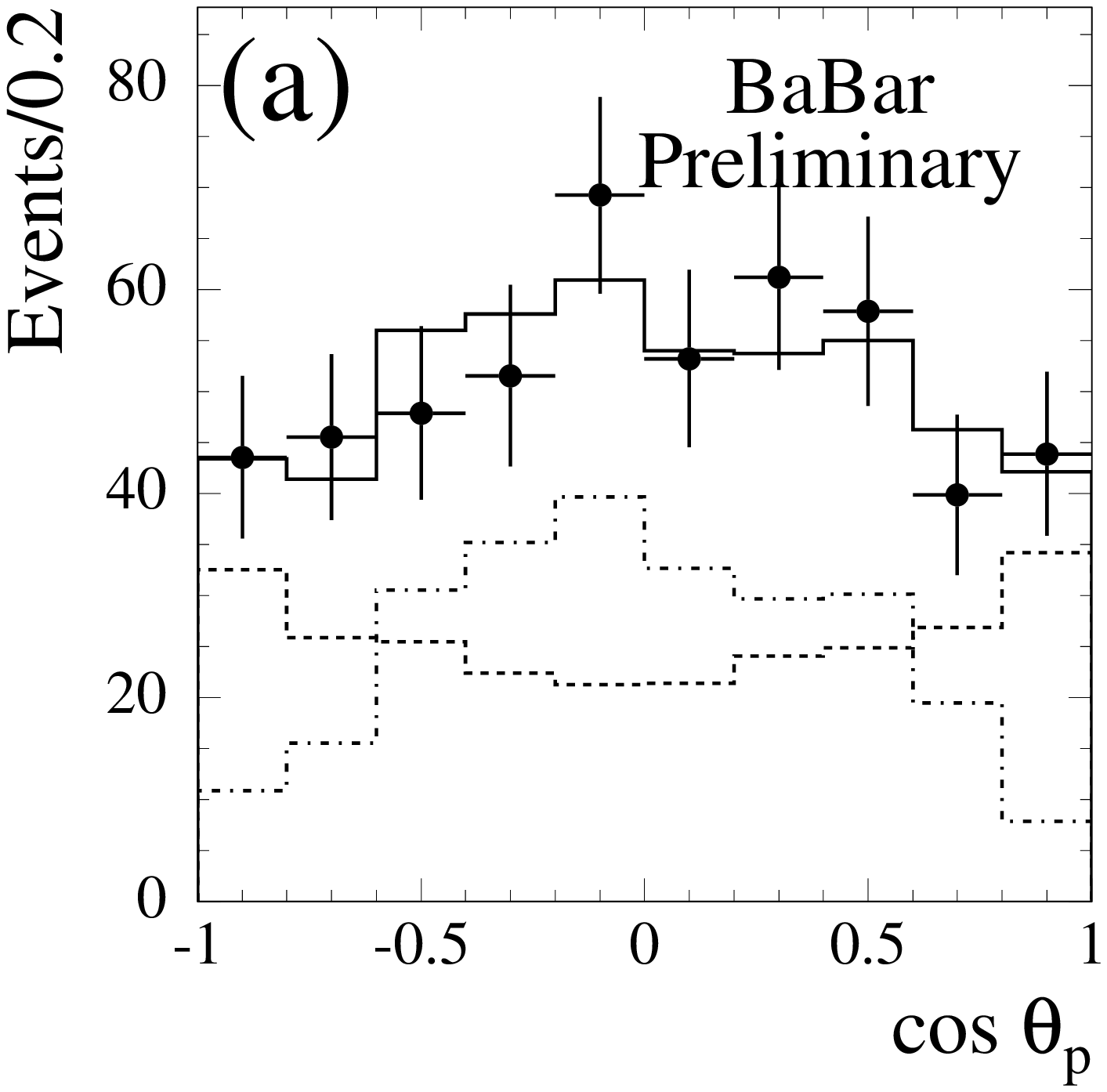} \\
\includegraphics[width=1.0\textwidth]{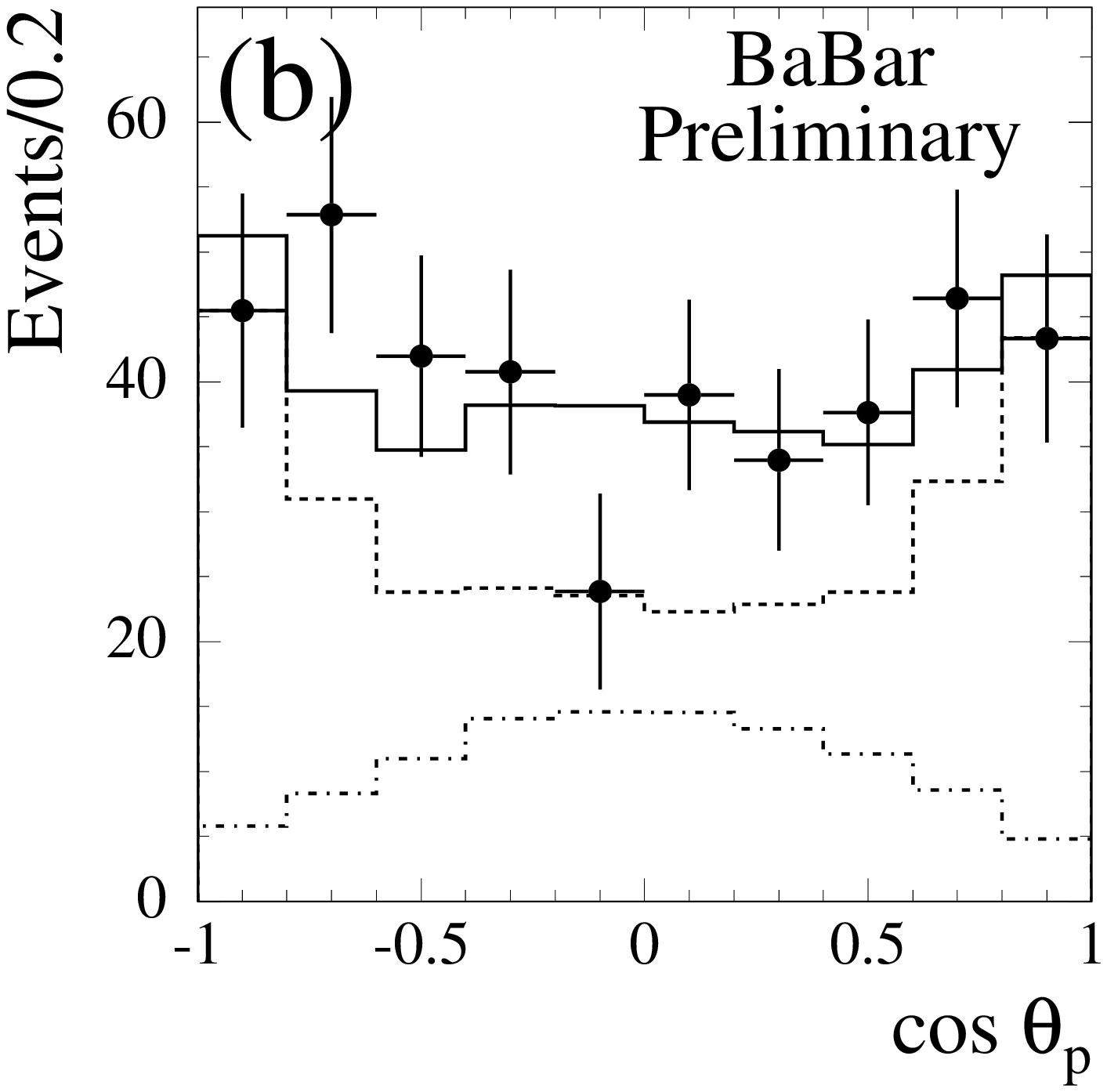}
\end{minipage}
\begin{minipage}{0.55\textwidth}
\includegraphics[width=1.0\textwidth]{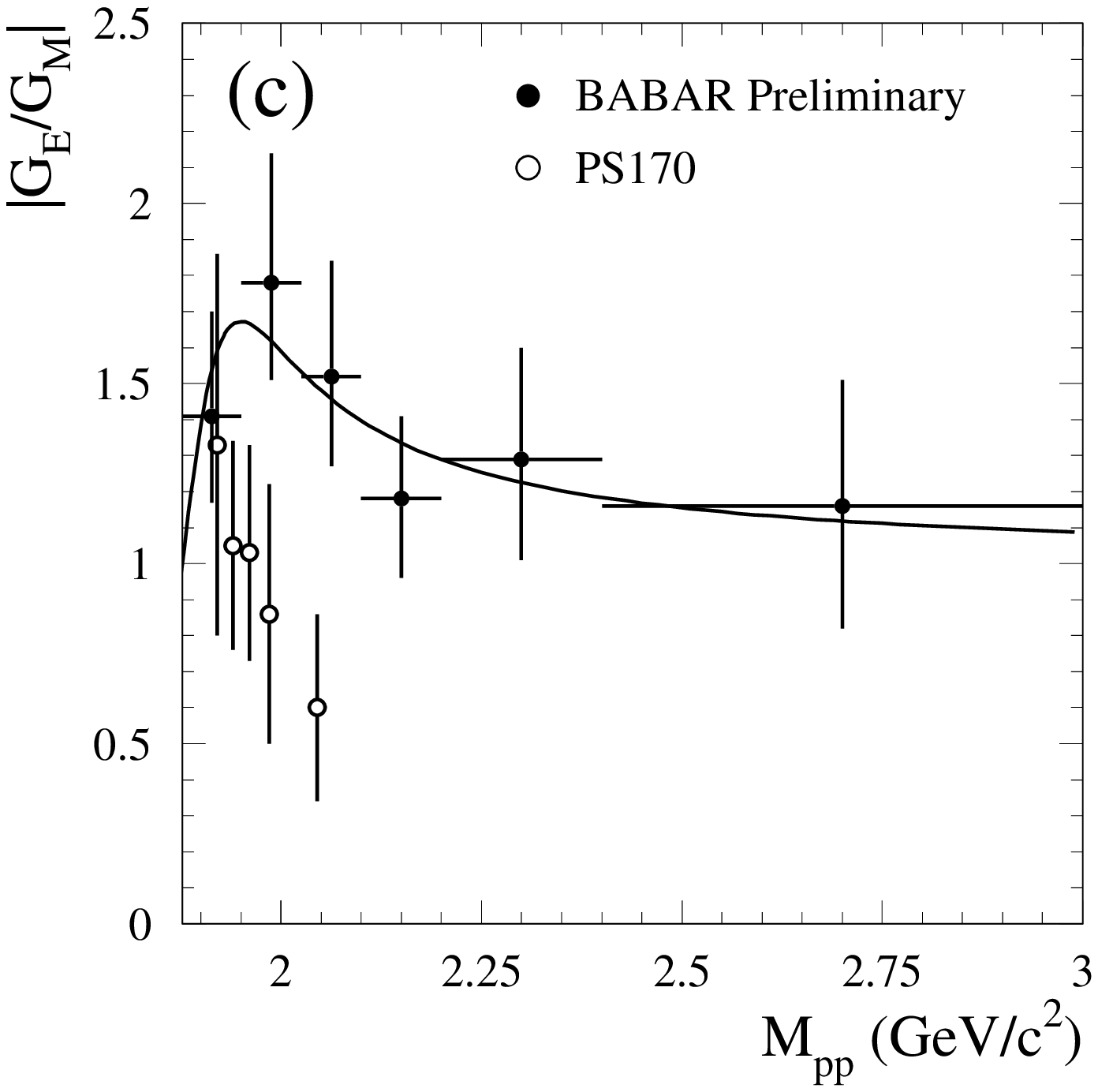}
\end{minipage}
\caption{$\cos \theta_p$ distributions for the mass bins $[1.877; 1.950] \gevcc$ (a)
and $[2.400; 3.000] \gevcc$ (b). The dashed (dot-dashed) curve shows the
contribution from the simulated $|G_M|^2$ ($|G_E|^2$) cross section term; the solid
line is the sum of the two terms and the points correspond to data. (c) Mass dependence
of the ratio $|G_E/G_M|$.}
\label{fig:gegm}
\end{figure}

\begin{equation}
\sigma_{\epem \to \ppbar}(s') = \frac{4\pi\alpha^2 C}{3s'}\sqrt{1 - \frac{2m_p^2}{s'}}
\left( |G_M(s')|^2 + \frac{2m_p^2}{s'} |G_E(s')|^2\right),
\end{equation}

\noindent where $m_p$ is the proton mass and $G_E$ and $G_M$ are respectively the 
electric and magnetic form factors of the proton. The factor $C$ accounts
for Coulomb interaction in the \ppbar system and has the 
effect of making the cross section non-zero at threshold.

The detection efficiency for the ISR process is about $17\%$, with no strong
dependence on either
the \ppbar invariant mass $m_{\ppbar}$ or the value of $|G_E/G_M|$.

The proton helicity angle $\theta_p$ in the \ppbar rest frame can be used to
separate the $|G_E|^2$ and $|G_M|^2$ terms. Their respective variations are approximately 
$\sin^2\theta_p$ and $1+\cos^2\theta_p$, with exact expressions obtained from 
Monte-Carlo simulation. By fitting the $\cos\theta_p$ distribution to a sum of the two terms
the ratio $|G_E/G_M|$ can be extracted. This is done separately in six bins 
of $m_{\ppbar}$. Preliminary results are shown in Fig.~\ref{fig:gegm}, and disagree significantly
with previous measurements from LEAR~\cite{lear}.

The cross section for $\epem \to \ppbar$ is measured and from it we obtain 
the ``effective form factor'' $G = \sqrt{|G_E|^2  + 2m_p^2/s' |G_M|^2}$. Preliminary values
are shown in Figs.~\ref{fig:xsplots}(a) and \ref{fig:xsplots}(b). Results are good agreement 
with existing data, and
cover the entire energy range from threshold to $4.5 \gev$ in a single measurement.
The data show a clear enhancement at threshold, 
already observed by LEAR, as well as hints of structures at $2.25$ and $\,3 \gevcc$.
Apart from these structures, the data agree well with a fit to the perturbative QCD 
expectation~\cite{chernyak,brodsky} $G(m_{\ppbar}) = A/m_{\ppbar}^4\log(m_{\ppbar}^2/\Lambda^2)$,
 where $A$ and $\Lambda$ are constants determined from the fit, which is shown as the dashed curve in
Fig.~\ref{fig:xsplots}(a).

\section{Final States With Six Hadrons}

\begin{figure}
\begin{minipage}{0.45\textwidth}
\includegraphics[width=1.0\textwidth]{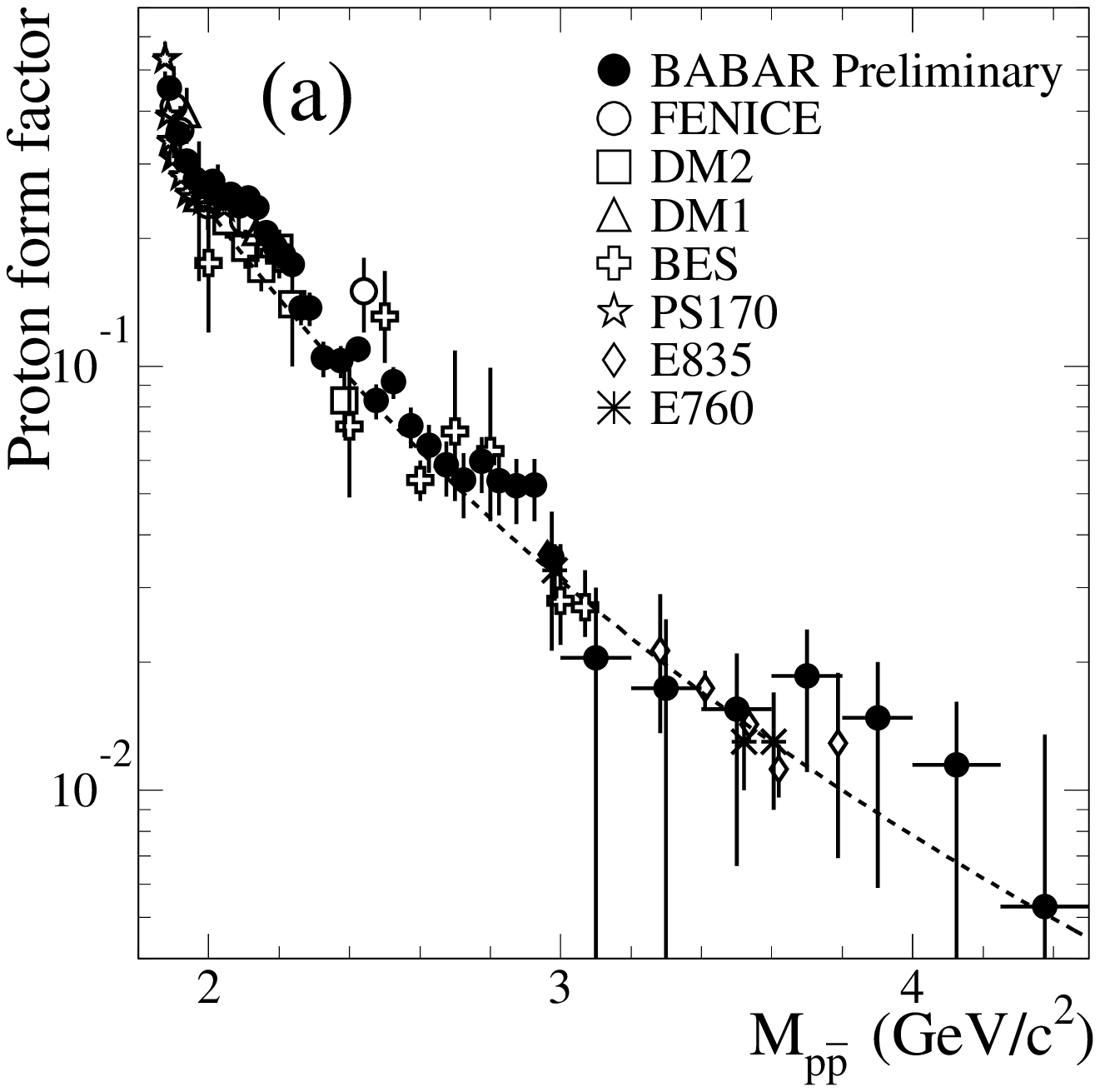}
\includegraphics[width=1.0\textwidth]{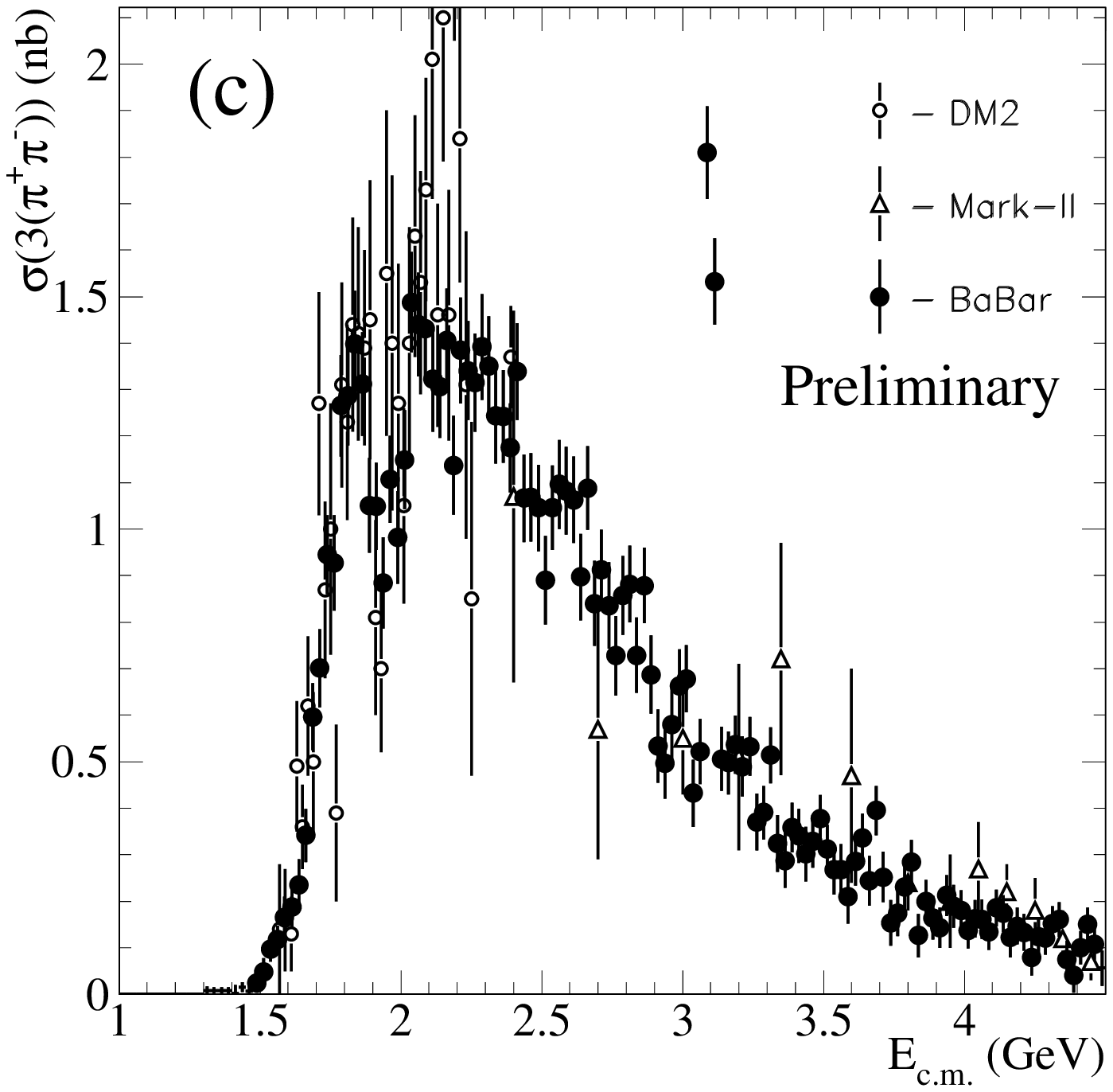}
\includegraphics[width=1.0\textwidth]{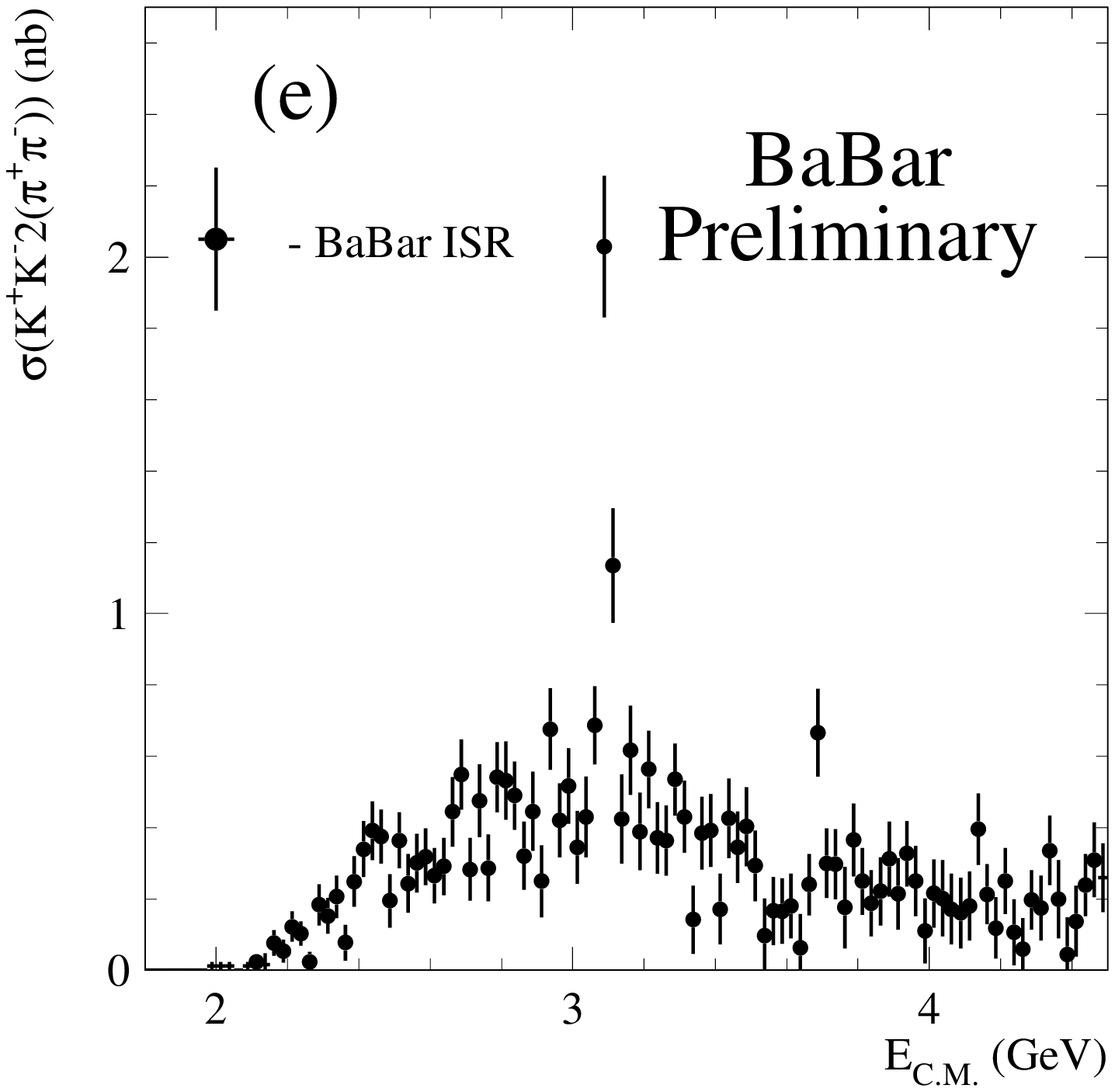}
\end{minipage}
\begin{minipage}{0.45\textwidth}
\includegraphics[width=1.0\textwidth]{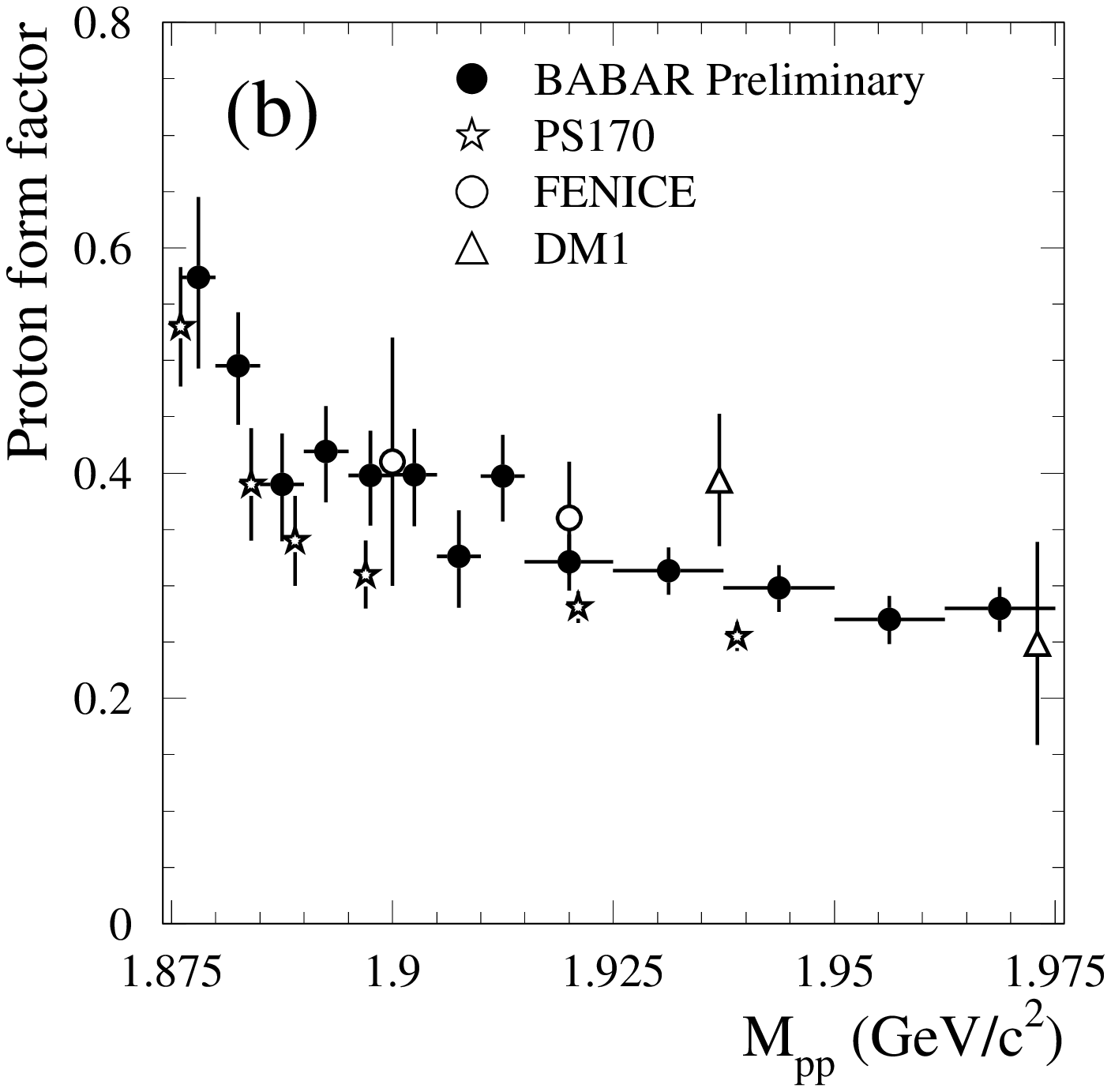}
\includegraphics[width=1.0\textwidth]{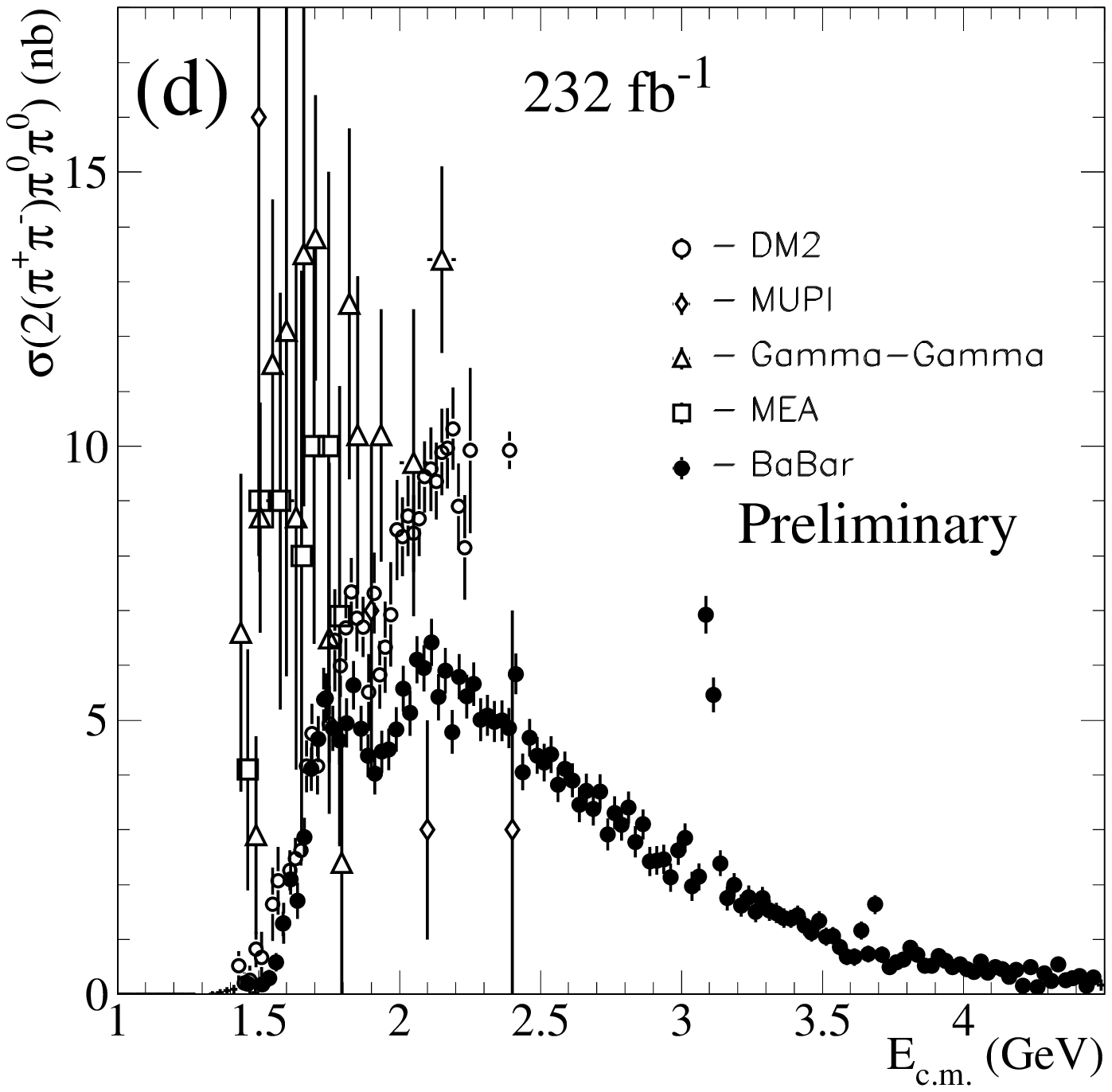}
\includegraphics[width=1.0\textwidth]{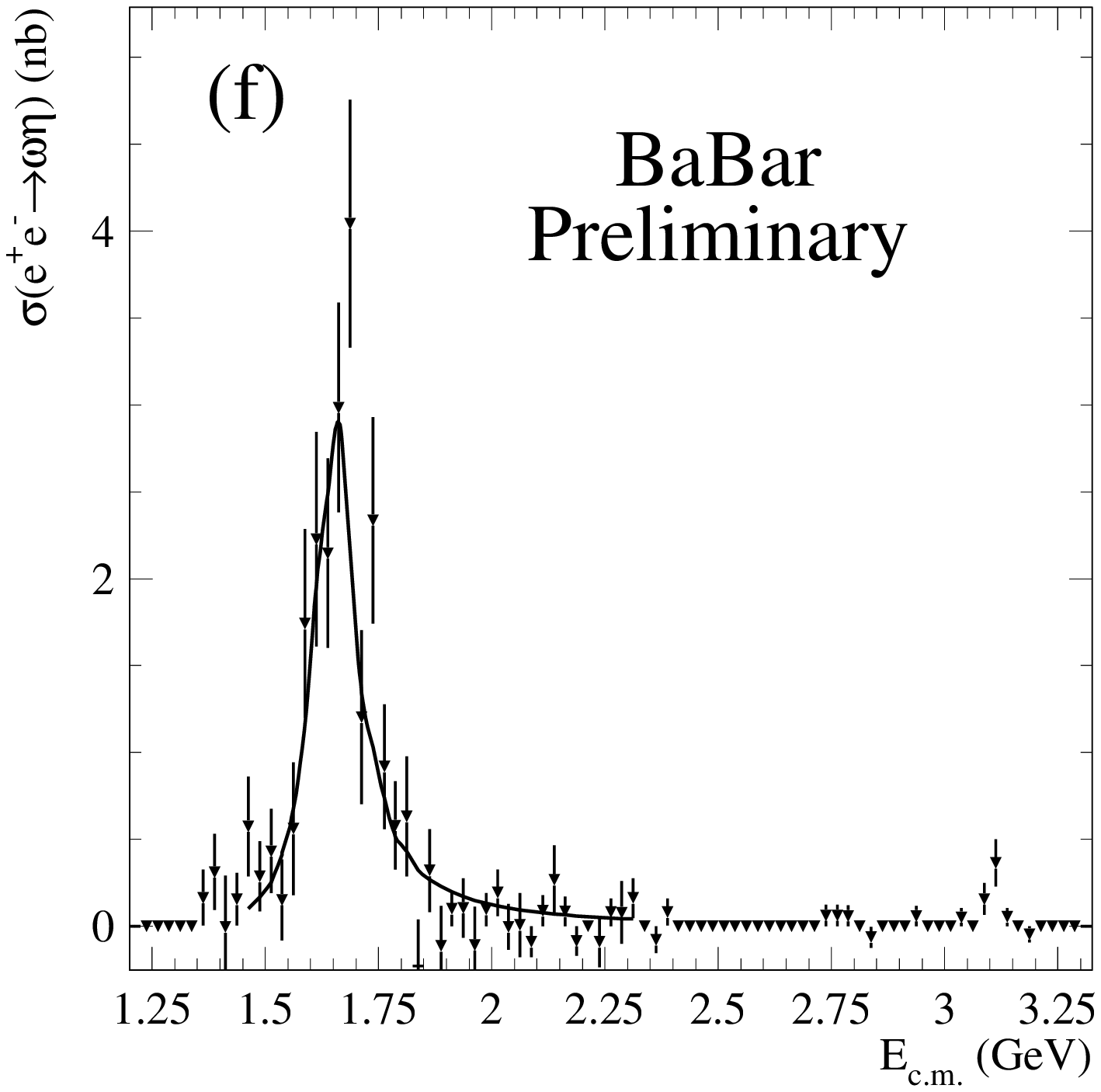}
\end{minipage}
\caption{(a) Effective proton form factor versus $m_{\ppbar}$; (b) 
Close-up view of the threshold region. Cross section distributions
versus $6h$ invariant mass for (c) $\epem \to 3(\pi^+\pi^-)$,
(d) $2(\pi^+\pi^-)2\pi^0$, (e) $K^+K^-2(\pi^+\pi^-)$,
and (f) $\epem \to \omega(\pi^+\pi^-\pi^0) \eta(\pi^+\pi^-\pi^0)$.}
\label{fig:xsplots}
\end{figure}

As an extension of the recent ISR measurements of the cross section of final states with 
four pions or kaons~\cite{fourh}, preliminary results have been obtained for 
the six-hadron processes $\epem \to 
3(\pi^+\pi^-)$, $2(\pi^+\pi^-)2\pi^0$ and $K^+K^-2(\pi^+\pi^-)$. The corresponding cross section
distributions are shown in Figs.~\ref{fig:xsplots}(c), \ref{fig:xsplots}(d) 
and \ref{fig:xsplots}(e) respectively. In the $2(\pi^+\pi^-)2\pi^0$ mode the data disagree strongly
with the DM2 results above $1.8 \gev$. Apart from this,
the data agrees well with existing results, and is considerably more precise.
The cross section for $K^+K^-2(\pi^+\pi^-)$ is measured for the first time. Again,
the entire energy range from threshold to $4.5 \gev$ is covered in a single experiment.

A clear dip is visible at about $1.9 \gev$ in the $6\pi$ modes.
A similar feature was already seen by DM2 in $\epem \to 3(\pi^+\pi^-)$ and
by FOCUS~\cite{focus} in the diffractive photoproduction of six charged pions. The
cross section distributions are fitted using the parametrization used by FOCUS, 
a sum of a Breit-Wigner resonance shape and a Jacob-Slansky continuum~\cite{js}. For the
$3(\pi^+\pi^-)$ ($2(\pi^+\pi^-)2\pi^0$) mode, we obtain values of $1880 \pm 30 \mev$
($1860 \pm 20 \mev$) for the resonance peak, $130 \pm 30 \mev$
($160 \pm 20 \mev$) for the resonance width  and $21 \pm 14 \;\mbox{deg}$
($-3 \pm 15 \;\mbox{deg}$) for phase shift between the resonance and continuum.
The width values differ significantly from the FOCUS result of $29 \pm 14 \mev$.

In the $3(\pi^+\pi^-)$ channel the resonance structure is surprisingly simple, being
well-described by a Monte-Carlo simulation featuring a single $\rho^0 \to \pi^+\pi^-$
resonance and the other four pions distributed according to phase space.
However rich resonant structures are observed in  the $2(\pi^+\pi^-)2\pi^0$
channel, with signals for $\,\rho^0 \to \pi^+\pi^-$, $\,\rho^+ \to \pi^+\pi^0$,
$\,f_0 \to \pi^0\pi^0$ and $\,f_0(1270) \to \pi^+\pi^-$ in the $2\pi$ combinations, 
and signals for $\omega$ and $\eta$ in $\pi^+\pi^-\pi^0$. A signal 
for $\epem \to \omega(\pi^+\pi^-\pi^0) \eta(\pi^+\pi^-\pi^0)$ is also seen, corresponding
to a resonant structure in the $6\pi$ invariant mass, as shown in Fig.~\ref{fig:xsplots}(f).
Fitting the distribution to a Breit-Wigner shape gives a peak position
of $1645 \pm 8 \mev$ and a width of $114 \pm 14 \mev$, which could correspond to the
$\omega''$ state reported in our previous analysis of the $\pi^+\pi^-\pi^0$ final state~\cite{threepi}.
Apart from this structure, which is unique to the 
$2(\pi^+\pi^-)2\pi^0$ mode, the ratio of the $2(\pi^+\pi^-)2\pi^0$ and $3(\pi^+\pi^-)$
cross sections is remarkably constant over the entire energy range, at a value of
$3.98 \pm 0.06 \mbox{(stat.)} \pm 0.41 \mbox{(syst.)}$.
In the $K^+K^-2(\pi^+\pi^-)$ final state, a signal for $\phi \to K^+K^-$ is seen, 
with a significant contribution from $J/\psi \to \phi \pi^+\pi^-\pi^+\pi^-$. Signals
for $K^{*0} \to K^+ \pi^-$ are also present.
Finally, the observed $J/\psi$ and $\psi(2S)$ signals can be used to extract branching 
fraction measurements which, in many cases, considerably improve over current 
world averages. These results are summarized in Table~\ref{tab:psi}.

\begin{table}[t]
\begin{tabular}{|l|c|c|}
\hline
Mode & Branching fraction (this work) & Branching Fraction (PDG)\\
\hline
$J/\psi \to 3(\pi^+\pi^-)$ & $(4.40 \pm 0.29 \pm 0.29) \times 10^{-3}$& $(4.0 \pm 2.0) \times 10^{-3}$\\
$J/\psi \to 2(\pi^+\pi^-)2\pi^0$ & $(1.65 \pm 0.10 \pm 0.18) \times 10^{-2}$& n/a\\
$J/\psi \to K^+K^-2(\pi^+\pi^-)$ &$(5.09 \pm 0.42 \pm 0.35) \times 10^{-3}$& $(3.1 \pm 1.3) \times 10^{-3}$\\
$\psi(2S) \to 2(\pi^+\pi^-)2\pi^0$ & $(5.3 \pm 1.6 \pm 0.6) \times 10^{-3}$& n/a\\
$\psi(2S) \to K^+K^-2(\pi^+\pi^-)$ & $(2.1 \pm 1.0 \pm 0.2) \times 10^{-3}$& n/a\\
\hline
\end{tabular}
\caption{Summary of $J/\psi$ and $\psi(2S)$ branching fraction obtained from six-hadron final states}
\label{tab:psi}
\end{table}

\section{Conclusion}

The program to measure low-energy hadronic cross sections using
ISR is well underway at BaBar. In addition to the published results for
$\,\pi^+\pi^-\pi^0$, $2(\pi^+\pi^-)$, $\,K^+K^-\pi^+\pi^-$ and $\,2(K^+K^-)$, 
we now have preliminary results for the $\,3(\pi^+\pi^-)$, $\,2(\pi^+\pi^-)2\pi^0$,
$\,K^+K^-2(\pi^+\pi^-)$ and \ppbar modes, with analyses 
ongoing for the final states $\,\pi^+\pi^-$, $\,K^+K^-$, $\,\pi^+\pi^-\pi^0\pi^0$, $\,\eta$, 
$\,\eta'$, and $\,D^{(*)}\bar{D}^{(*)}$.

\section{Acknowledgments}

The author is grateful for the extraordinary contributions of the PEP-II colleagues in achieving the excellent
luminosity and machine conditions that have made this work possible. This work is supported by
DOE
and NSF (USA),
NSERC (Canada),
IHEP (China),
CEA and
CNRS-IN2P3
(France),
BMBF and DFG
(Germany),
INFN (Italy),
FOM (The Netherlands),
NFR (Norway),
MIST (Russia), and
PPARC (United Kingdom).
Individuals have received support from CONACyT (Mexico), A.~P.~Sloan Foundation,
Research Corporation,
and Alexander von Humboldt Foundation.

\end{document}